\documentclass[
 reprint,
 superscriptaddress,
 amsmath,amssymb,
 aps,
pra,
showkeys
]{revtex4-2}

\usepackage{soul}
\usepackage{float}
\usepackage{multibib}
\usepackage{hyperref}
\usepackage{xcolor}
\hypersetup{
    colorlinks,
    linkcolor={blue!50!blue},
    citecolor={blue!50!blue},
    urlcolor={blue!80!black}
}
\usepackage{siunitx}
\usepackage{import}
\usepackage{placeins}
\usepackage{xcolor}
\usepackage{lmodern}
\usepackage[braket, qm]{qcircuit}
\usepackage{graphicx}% Include figure files
\usepackage{dcolumn}% Align table columns on decimal point
\usepackage{bm}% bold math
\usepackage{comment}
\usepackage[english]{babel}
\usepackage{blindtext}
\usepackage{physics}  %for ket,bra,expectation value
\usepackage{multirow}
\usepackage{array}

\begin{document}

\title{Theory of the CH$_\text{S}^-$ Defect in MoS$_2$ Confirming the Origin of the Quantum Emission}  

%Microscopic Origin of CH$_\mathrm{S}^{-}$ Bound-Exciton Emission in MoS$_2$

\author{Chanaprom Cholsuk}
\email{chanaprom.cholsuk@tum.de}
\affiliation{Department of Computer Engineering, TUM School of Computation, Information and Technology, Technical University of Munich, 80333 Munich, Germany}
\affiliation{Munich Center for Quantum Science and Technology (MCQST), 80799 Munich, Germany}

\author{Tobias Vogl}%
\email{tobias.vogl@tum.de}
\affiliation{Department of Computer Engineering, TUM School of Computation, Information and Technology, Technical University of Munich, 80333 Munich, Germany}
\affiliation{Munich Center for Quantum Science and Technology (MCQST), 80799 Munich, Germany}

\date{\today}% It is always \today, today,

\begin{abstract}
Establishing the microscopic origin of localized emission in two-dimensional materials through consistent agreement between experiment and theory remains challenging. Recent STM/STS-STML measurements assigned an emission near 1.8~eV in monolayer MoS$_2$/hBN(2L)/graphene to the negatively charged CH$_\mathrm{S}^{-}$ defect. However, a comprehensive theoretical description of this emitter is still lacking. Here, we develop such a description and systematically investigate the photophysical properties of CH$_\mathrm{S}^{-}$ in monolayer MoS$_2$, two bilayer configurations, and a MoS$_2$/hBN heterostructure. The defect retains a similar local geometry and favors a singlet ground state in all four environments, whereas its electronic transition, zero-phonon line, electron--phonon coupling, radiative lifetime, and transition-dipole orientation depend strongly on the surrounding layers. The isolated-monolayer model exhibits substantial structural reorganization and produces a broad, dominant phonon sideband that is inconsistent with experiment. Adding an adjacent layer, either MoS$_2$ or hBN, substantially suppresses the phonon sideband, with only a minor dependence on the defect position in bilayer MoS$_2$. In particular, the MoS$_2$/hBN model yields a weak phonon sideband and closely reproduces the experimental photoluminescence lineshape. Our results thus provide the first theoretical agreement of CH$_\mathrm{S}^{-}$ emission and further reveal the role of adjacent layers in determining the electronic, vibronic, and optical properties of the quantum emitters in two-dimensional materials.
\end{abstract}


\maketitle
% 1500 and 4000 words

\section{Introduction}
Single-photon sources (SPSs) in two-dimensional (2D) materials have attracted considerable attention because of their efficient photon extraction \cite{10.1021/acsnano.3c08940}, compatibility with integrated photonic architectures \cite{10.1088/1361-6463/aa7839,10.1002/qute.2023003433}, radiation tolerance \cite{10.1038/s41467-019-09219-5}, and reduced density of surface dangling bonds, which can help suppress decoherence channels \cite{10.1038/s41524-025-01859-0}. Among 2D materials, defects in hexagonal boron nitride \cite{10.1038/nnano.2015.242,10.1063/5.0147560,10.1039/D5TC02805A,
10.1021/acs.jpcc.4c03404} and transition-metal dichalcogenides (TMDs) constitute a particularly promising platform for localized quantum emission. For TMDs, single-photon emission was first extensively reported in monolayer WSe$_2$ \cite{10.1364/OPTICA.2.000347,10.1038/ncomms15093,10.1038/s41699-020-0136-0,10.1038/s41467-021-23709-5,10.1038/nnano.2015.75}, and has subsequently been observed in a range of TMDs, including MoS$_2$ \cite{10.1038/s41467-019-10632-z,10.1021/acsnano.2c09209}, MoSe$_2$ \cite{10.1038/nnano.2013.277,10.1021/acs.nanolett.0c04282}, WS$_2$ \cite{10.1038/ncomms15093,10.1038/s41467-021-27585-x,10.1126/sciadv.abb5988}, and MoTe$_2$ \cite{10.3390/nano13091501,arxiv:2508.20743}. In these systems, localized emission has been attributed to several mechanisms, including strain-induced band bending, defect-induced localized excitonic states, and the combined influence of atomic defects, strain, and local electrostatic potentials \cite{10.1021/acsnano.9b02316,10.1021/acs.nanolett.5b03312,10.1038/s41467-021-23709-5,10.1088/2053-1583/ab15fe}.\\
\begin{figure*}[ht!]
    \centering
    \includegraphics[width=1\linewidth]{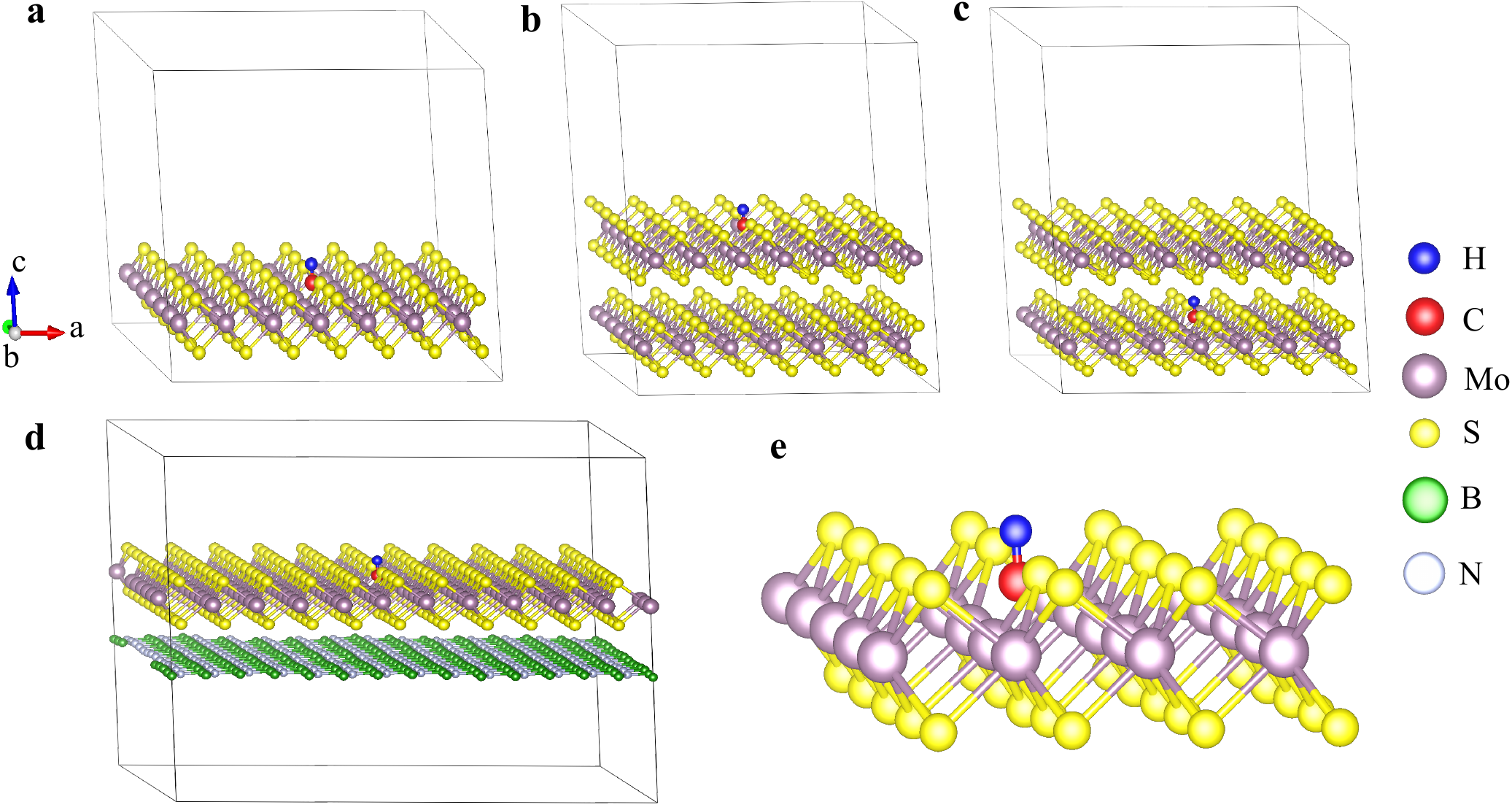}
    \caption{\textbf{Optimized geometries of CH$_\mathrm{S}^{-}$ in each system.} 
    \textbf{a} MoS$_2$ monolayer; \textbf{b} MoS$_2$ bilayer with CH$_\mathrm{S}^{-}$ at the top layer;  \textbf{c} MoS$_2$ bilayer with CH$_\mathrm{S}^{-}$ at the middle layer; \textbf{d} MoS$_2$/hBN heterostructure; and \textbf{e} Zoom-in CH$_\mathrm{S}^{-}$ defect on the MoS$_2$ layer.}
    \label{fig:structure}
\end{figure*}
\indent Despite substantial progress, assigning a specific microscopic origin to a given optical emitter remains a central challenge.~Such microscopic identification is essential because the optical energy, spin configuration, electron--phonon coupling, and quantum efficiency are all controlled by the atomic defect structure. Establishing the emitter origin therefore provides the basis for rationally tuning its optical properties and for deterministic defect engineering, especially for quantum technology applications \cite{10.3390/nano12142427,10.1002/adom.202402508, 10.1063/5.0188597}. In CVD-grown WSe$_2$, for example, combined scanning tunneling microscopy, scanning transmission electron microscopy, and GW--BSE calculations suggested that oxygen interstitials, together with strain, are likely responsible for localized excitons for its SPS \cite{10.1021/acsnano.9b02316}. Carbon-related defects have also emerged as important candidates in TMDs. In WSe$_2$ and WS$_2$, chalcogen-site carbon impurities, C$_\mathrm{Se}$ and C$_\mathrm{S}$, have been proposed as promising emitters. These defects can be generated through STM-induced hydrogen desorption from carbon--hydrogen complexes, and their optical properties have subsequently been investigated theoretically \cite{10.1038/s41467-021-27585-x,10.1038/s41467-022-28876-7}. Together, these studies demonstrate that reliable emitter assignment requires a direct connection between experiment and theory.\\
\indent Recently, scanning tunneling microscopy/spectroscopy combined with STM-induced luminescence (STM/STS-STML) provided such an atomistic correlation for individual defects in monolayer MoS$_2$ \cite{arXiv:2510.15676}. In that experiment, monolayer MoS$_2$ was electronically decoupled from graphene by bilayer hBN, enabling sharp STML emission from pristine excitons and trions. They found that O$_\mathrm{S}$ and V$_\mathrm{S}^{-}$ mainly suppress the pristine MoS$_2$ exciton/trion emission, whereas negatively charged CH$_\mathrm{S}^{-}$ defects produce a distinct defect-bound exciton emission approximately 200~meV below the pristine MoS$_2$ exciton. This emission was interpreted as an acceptor-bound exciton, A$^{-}$X, arising from a band-to-defect optical transition whose energy follows the local band bending around the charged defect. However, a theoretical understanding of the CH$_\mathrm{S}^{-}$ defect and its theoretical optical properties in MoS$_2$ is still lacking.\\
\indent In this work, we herein employ density functional theory (DFT) to investigate the structural, electronic, vibronic, and optical properties of the CH$_\mathrm{S}^{-}$ defect. As the experiment was performed in a MoS$_2$/hBN(2L)/graphene heterostructure, we first investigate its intrinsic properties in monolayer MoS$_2$ and then examine how interlayer coupling and dielectric screening modify its photophysics in bilayer MoS$_2$ and a MoS$_2$/hBN heterostructure. Although an explicit treatment of the complete four-layer experimental structure is computationally prohibitive, these models capture the principal effects introduced by additional MoS$_2$ and hBN layers and clarify how the surrounding layered environment influences the optical response of CH$_\mathrm{S}^{-}$. For each system, we determine the relaxed defect geometry, defect-induced electronic states, spin configuration, zero-phonon line (ZPL), Huang--Rhys (HR) factor, Debye--Waller (DW) factor, photoluminescence (PL) spectrum, radiative lifetime, and transition-dipole orientation.  Our results establish CH$_\mathrm{S}^{-}$ as a well-defined atomic candidate for localized excitonic emission in MoS$_2$ and provide microscopic guidance for further identifying and engineering defect-bound optical transitions.

\section{Results}
To characterize the photophysical properties of the CH$_\mathrm{S}^{-}$ defect in MoS$_2$, we first determine its relaxed atomic structure and ground-state spin configuration. We then identify the defect-induced electronic states and the electronic transition responsible for the predicted optical emission. Next, we quantify its optical and vibronic properties, including the ZPL, HR and DW factors, PL spectrum, radiative lifetime, and transition-dipole orientation. Finally, we investigate how interlayer coupling modifies these properties in monolayer MoS$_2$, bilayer MoS$_2$, and the MoS$_2$/hBN heterostructure.

\begin{table*}[t]
\centering
\caption{Optimized structural parameters of CH$_\mathrm{S}^{-}$ in four different hosts where $\Delta z_{\mathrm{C}}$ denotes the inward displacement of C from the substituted S plane.}
\label{tab:structure}
\footnotesize
\setlength{\tabcolsep}{7pt}
\renewcommand{\arraystretch}{1.08}
\begin{tabular}{lcccc}
\hline
Parameter & Monolayer & Top-bilayer & Midle-bilayer & Heterostructure\\
\hline
Lattice constant $a$  (\AA) & 3.160 & 3.160 & 3.160 & 3.160 \\
C--H bond length (\AA) & 1.088 & 1.088 & 1.088 & 1.088 \\
Nearby $\langle d_{\mathrm{Mo-C}}\rangle$ (\AA) & 2.114 & 2.114 & 2.114 & 2.113 \\
Nearby $\langle d_{\mathrm{Mo-S}}\rangle$ (\AA) & 2.397 & 2.394 & 2.395& 2.394 \\
$\Delta z_{\mathrm{C}}$ (\AA) & 0.453 & 0.448 & 0.442 & 0.450 \\
C--H tilt from $z$ axis ($^\circ$) & 0.237 & 0.111 & 0.039 & 0.148 \\
Layer separation (\AA) & -- & 3.210 & 3.194 & 3.616 \\
\hline
\end{tabular}
\end{table*}

\begin{figure*}[ht!]
    \centering
    \includegraphics[width=1\linewidth]{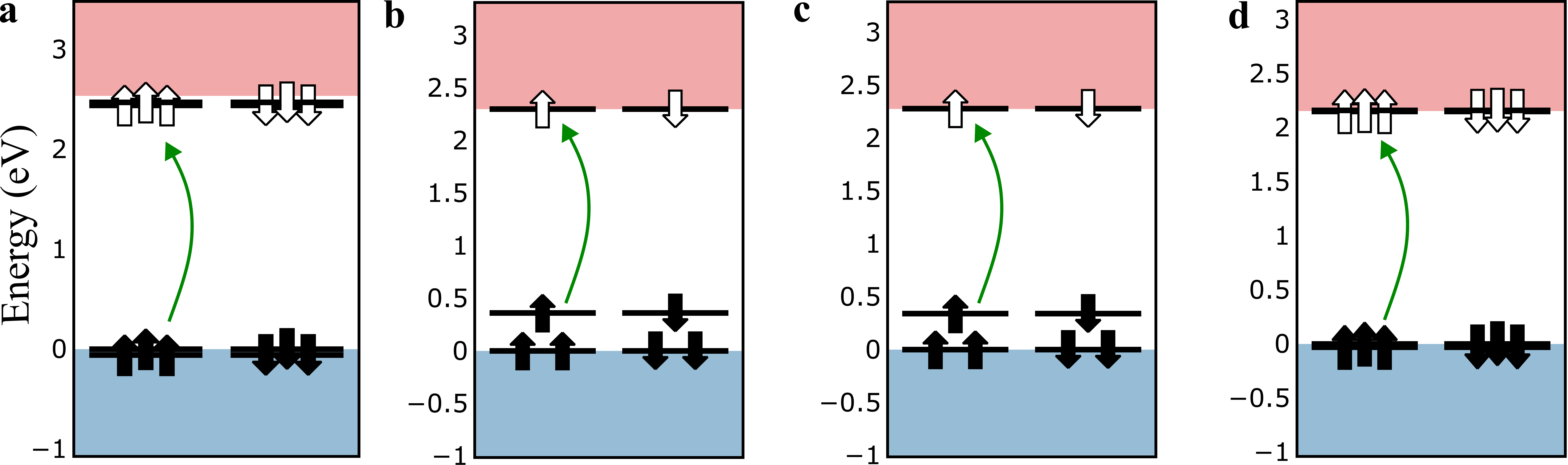}
    \caption{\textbf{Kohn--Sham electronic transitions at the $\Gamma$ point for CH$_\mathrm{S}^{-}$ in MoS$_2$.} 
\textbf{a} Monolayer MoS$_2$. 
\textbf{b} Bilayer MoS$_2$ with CH$_\mathrm{S}^{-}$ at the outer/top S site. 
\textbf{c} Bilayer MoS$_2$ with CH$_\mathrm{S}^{-}$ at the inner/middle S site. \textbf{d} Heterostructure MoS$_2$/hBN with CH$_\mathrm{S}^{-}$. 
Filled and unfilled arrows denote occupied and unoccupied Kohn--Sham states, respectively, while upward and downward arrows indicate spin-up and spin-down channels. 
The green arrow marks the electronic transition used to calculate the optical properties.}
    \label{fig:electronic}
\end{figure*}

\subsection{Optimizing geometry}
The CH$_\mathrm{S}^{-}$ defect is positioned near the center of the MoS$_2$ layer in each model, as shown in Fig.~\ref{fig:structure}. The atomic structures are subsequently optimized within DFT using the HSE06 hybrid functional. For monolayer (148 atoms) and bilayer MoS$_2$ (295 atoms), we employ a $7\times7\times1$ supercell to minimize finite-size effects arising from interactions between periodic defect images. For the MoS$_2$/hBN heterostructure, we construct a commensurate supercell consisting of an $11\times11\times1$ MoS$_2$ layer and a $14\times14\times1$ hBN layer, containing 756 atoms in total. This configuration minimizes both finite-size effects and the strain associated with the lattice mismatch, leaving a residual strain of approximately $0.69\%$ in the hBN layer. This represents the smallest residual strain that is computationally feasible within the present study. The next-larger commensurate structure, comprising a $15\times15\times1$ MoS$_2$ layer and a $19\times19\times1$ hBN layer, would reduce the residual strain to $0.21\%$ but would contain 1398 atoms, which is computationally challenging for the HSE06-level calculations. See Sec.~\ref{sec:method-DFT} for further computational details.\\
\indent As shown in Fig.~\ref{fig:structure} and summarized in Tab.~\ref{tab:structure}, the CH$_\mathrm{S}^{-}$ defect adopts a nearly identical local geometry in all four host environments. The C--H bond length remains fixed at $1.088$~\AA, while the average Mo--C distance varies by only $0.001$~\AA. Moreover, the C atom undergoes a similar inward displacement of $0.442$--$0.453$~\AA\ from the substituted S plane, and the C--H bond remains almost parallel to the $z$ axis. These results indicate that the relaxed atomic structure of CH$_\mathrm{S}^{-}$ is only weakly affected by the presence and identity of neighboring layers. Consequently, any substantial differences in its electronic or optical properties are unlikely to originate from changes in the local defect geometry alone.

\begin{figure*}
    \centering
    \includegraphics[width=0.75\linewidth]{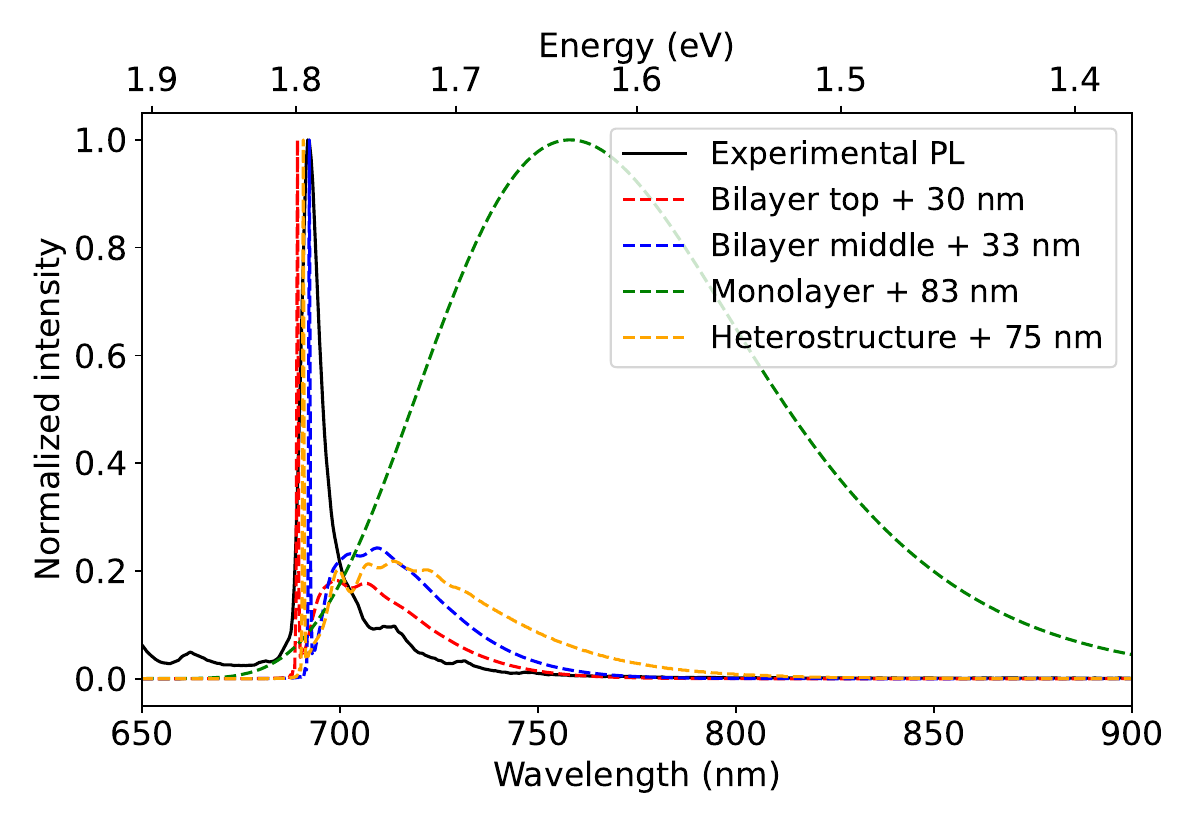}
    \caption{\textbf{Comparison of the calculated and experimental PL spectra of CH$_\mathrm{S}^{-}$.} Calculated spectra are shown for CH$_\mathrm{S}^{-}$ in monolayer MoS$_2$, the top- and middle-bilayer configurations, and the MoS$_2$/hBN heterostructure, together with the experimental spectrum. The spectra are normalized to their respective maximum intensities, and the calculated ZPL positions are aligned with the experimental emission peak to compare the PL lineshapes and relative PSB contributions independently of the absolute transition energies. Experimental data is adapted with permission from Ref.~\cite{arXiv:2510.15676}.}
    \label{fig:PL}
\end{figure*}

\begin{figure*}
    \centering
    \includegraphics[width=1\linewidth]{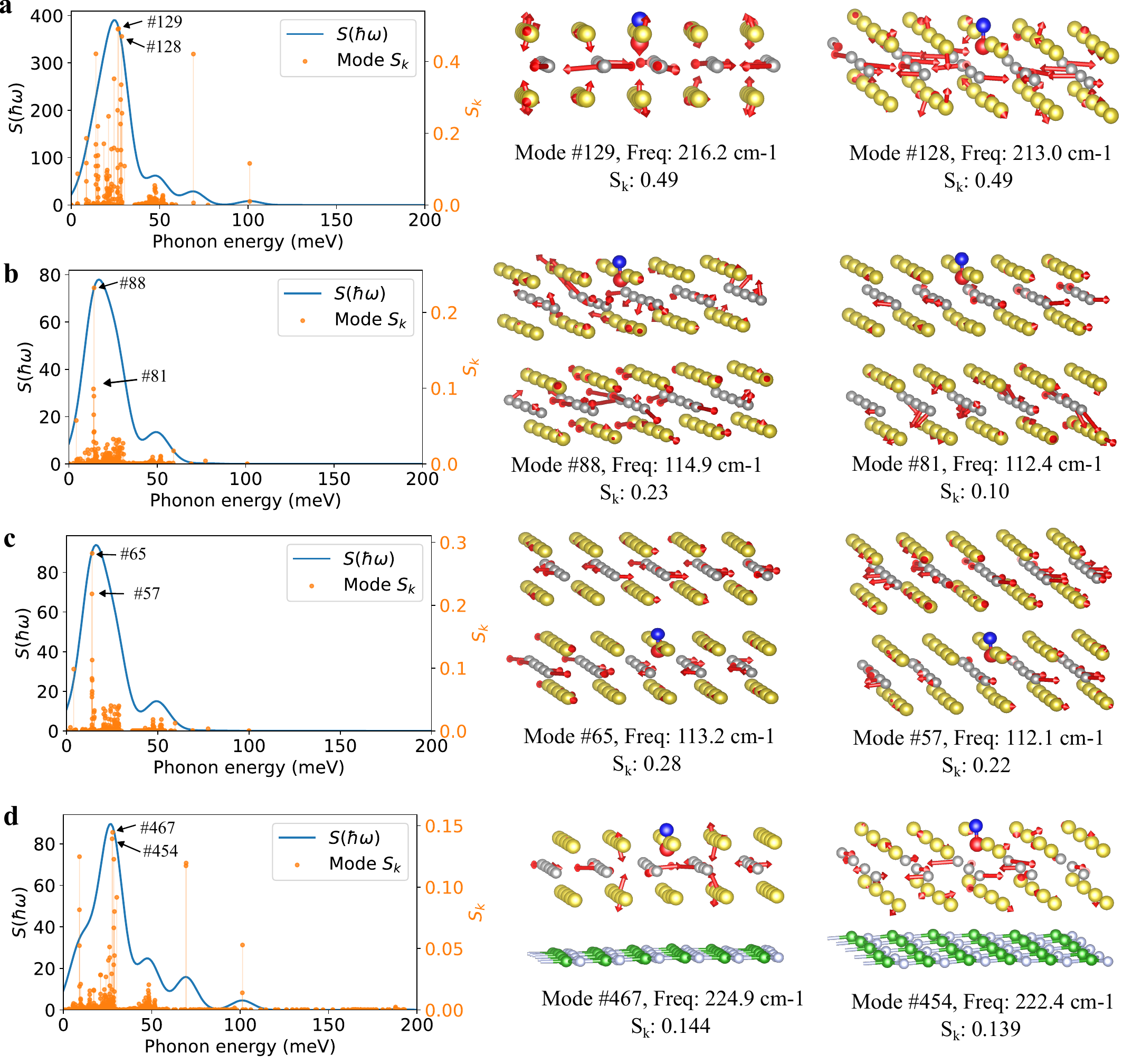}
    \caption{\textbf{Simulated phonon spectral density of CH$_\mathrm{S}^{-}$ and the most two dominant phonon modes. }
\textbf{a} Monolayer MoS$_2$.
\textbf{b,c} Bilayer MoS$_2$ with CH$_\mathrm{S}^{-}$ at the outer/top S site and inner/middle S site, respectively. 
 \textbf{d} MoS$_2$/hBN heterostructure with CH$_\mathrm{S}^{-}$ defect on the MoS$_2$ layer. The red arrow represents the vibrational direction. }
 \label{fig:PSB}
\end{figure*}

\subsection{Electronic transition and optical transition}
We next examine the spin multiplicity and electronic transitions of CH$_\mathrm{S}^{-}$ in the four host environments. We found that geometry optimizations initialized with different spin configurations consistently yield a singlet ground state for all the four systems. \\
\indent As shown in Fig.~\ref{fig:electronic}, the character of the relevant optical transition depends on the surrounding layered environment. In monolayer MoS$_2$ and the MoS$_2$/hBN heterostructure, the electronic transition of CH$_\mathrm{S}^{-}$ occurs predominantly between defect states close to the valence- and conduction-band edges. In contrast, when CH$_\mathrm{S}^{-}$ is located in either the upper or middle layer of the bilayer MoS$_2$, the transition originates from an isolated occupied defect state to the defect states close to the conduction band. This suggests that the presence of CH$_\mathrm{S}^{-}$ defect introduces defect states that participate in the optical transition, but their positions relative to the host bands and their contributions to the transition vary with the surrounding layers. We also found that all four systems possess a singlet ground state, so the spin-up and spin-down channels are energetically degenerate in the absence of spin--orbit-induced splitting. The corresponding spin-conserving transitions are therefore expected to yield identical ZPL energies. Accordingly, we calculate the ZPL using only the spin-up channel. \\
\indent The resulting ZPL energies are 2.14, 1.87, 1.88, and 2.01~eV for CH$_\mathrm{S}^{-}$ in monolayer MoS$_2$, the upper layer of bilayer MoS$_2$, the middle layer of bilayer MoS$_2$, and the MoS$_2$/hBN heterostructure, respectively, as also summarized in Tab.~\ref{tab:properties}. Relative to the experimentally reported emission peak of approximately 1.80~eV \cite{arXiv:2510.15676}, these values differ by 18.9\%, 3.9\%, 4.4\%, and 11.7\%, respectively. The variation among the calculated ZPL energies demonstrates that the surrounding layered environment appreciably influences the optical transition of CH$_\mathrm{S}^{-}$. We note that the calculated ZPL energies may also be affected by the finite supercell size. In our previous study of a defect in a TMD bilayer, increasing the supercell from $7\times7\times1$ to $9\times9\times1$ changed the ZPL by only 5~meV \cite{arxiv:2605.18501}. This small variation indicates that the $7\times7\times1$ supercell provides a well-converged ZPL value. Our calculated ZPL values can therefore be regarded as representative of the ZPLs predicted for these specific structural models within the present HSE06 framework.\\
\indent Although the two bilayer configurations yield ZPL energies closest to the experimental value, we rule out a bilayer MoS$_2$ environment as the origin of the observed transition because the STM/STS-STML measurements were performed on a specifically prepared monolayer MoS$_2$/hBN(2L)/graphene heterostructure. The atomic-layer composition of such samples can be determined with high precision, and a bilayer MoS$_2$ region would be directly distinguishable in the experiment. We therefore regard the close numerical agreement obtained for the bilayer models as coincidental rather than as evidence for a bilayer origin of the measured emission. Among the models considered here, the MoS$_2$/hBN heterostructure most closely represents the experimentally established local environment of CH$_\mathrm{S}^{-}$, despite its somewhat larger deviation from the measured emission peak. \\
\indent The remaining difference between the calculated heterostructure ZPL of 2.01~eV and the experimental value of approximately 1.80~eV may arise partly from the residual strain associated with the commensurate computational supercell. To examine the possible role of strain, we applied uniaxial and biaxial strain to monolayer MoS$_2$ containing CH$_\mathrm{S}^{-}$. Our strain-dependent results show that applying only $0.5\%$ uniaxial or biaxial strain can shift the predicted emission wavelength by approximately 0.1 eV or around 30~nm, even in monolayer MoS$_2$. This demonstrates the pronounced strain sensitivity of the CH$_\mathrm{S}^{-}$ optical transition. The residual lattice-mismatch strain in the heterostructure model may therefore contribute appreciably to the calculated ZPL. Additional differences may originate from the simplified representation of the experiment, particularly the omission of the second hBN layer and the graphene substrate. Consequently, the experimentally relevant ZPL may be shifted toward the measured value when the residual strain and the full layered environment are taken into account. Nevertheless, quantitatively establishing the contribution of strain will require an independent experimental determination of the local strain in the vicinity of the CH$_\mathrm{S}^{-}$ defect.

\subsection{Optical properties}
Although the calculated ZPLs show reasonable agreement with the experimental emission energy, comparison of a single optical property is generally insufficient for a reliable microscopic assignment and may lead to misidentification of the defect \cite{10.1021/acs.jpclett.3c01475}. A more comprehensive comparison requires the PL lineshape, PSB contribution, HR factor, and other vibronic properties. We therefore calculate the PL spectra using the full set of phonon modes. The underlying finite-displacement calculations comprise 888, 1770, and 4536 displaced structures for CH$_\mathrm{S}^{-}$ in monolayer MoS$_2$, bilayer MoS$_2$, and the MoS$_2$/hBN heterostructure, respectively. Further details of the PL simulations are provided in Sec.~\ref{sec:method-PL}.\\
\indent Fig.~\ref{fig:PL} compares the calculated PL spectra of the four configurations with the experimental spectrum \cite{arXiv:2510.15676}. For monolayer MoS$_2$, the calculated spectrum shows a broad and dominant PSB. This is reflected in its large HR factor of approximately 10. As shown in Fig.~\ref{fig:PSB}\textbf{a}, one of the two modes with the largest partial HR factors involves vibrations of the C--H defect together with the surrounding MoS$_2$ lattice, whereas the other is dominated by host-lattice vibrations. The participation of the defect in a strongly coupled mode indicates that the optical transition is accompanied by local structural relaxation. Indeed, the monolayer has a large configuration-coordinate displacement of $\Delta Q=1.99$ amu$^{1/2}$\AA. This reflects substantial structural reorganization between the ground and excited states. This large displacement is consistent with the strong electron--phonon coupling and broad PSB.\\
\indent For bilayer MoS$_2$, the top- and middle-bilayer configurations show similar PL spectra, suggesting that the defect position has only a modest effect on the overall PL lineshape. In both cases, the PSB is much weaker than in the monolayer, consistent with HR factors of 2.12 and 2.39, respectively. As shown in Fig.~\ref{fig:PSB}\textbf{b} and Fig.~\ref{fig:PSB}\textbf{c}, the two modes with the largest partial HR factors are dominated by vibrations of the MoS$_2$ host, without appreciable motion of the CH$_\mathrm{S}^{-}$ defect. The $Q$ values are also smaller than in the monolayer, with $\Delta Q=1.06$ and 1.13 amu$^{1/2}$\AA~ for the top- and middle-bilayer configurations, respectively. These values indicate less structural reorganization between the ground and excited states and are consistent with the smaller HR factors and weaker PSBs. \\
\indent For the MoS$_2$/hBN heterostructure, the PL spectrum likewise shows a relatively weak PSB, with an HR factor of 2.70. The two modes with the largest partial HR factors, shown in Fig.~\ref{fig:PSB}\textbf{d}, involve vibrations of the MoS$_2$ host, while neither the CH$_\mathrm{S}^{-}$ defect nor the hBN layer shows appreciable motion. The $Q$ value is $1.04$ amu$^{1/2}$\AA, indicating less structural reorganization than in the monolayer. Although this value is slightly smaller than those of the bilayer configurations, the total HR factor is slightly larger because it depends on the mode-resolved displacements and phonon frequencies, not on $\Delta Q$ alone.\\
\indent Overall, the monolayer calculation predicts a much broader and stronger PSB than observed experimentally. The isolated-monolayer model is therefore strongly disfavored as an explanation of the measured PL spectrum. Both bilayer configurations and the MoS$_2$/hBN heterostructure reproduce the comparatively weak PSB more closely. Although all three models contain two layers, their dominant phonon modes differ: both MoS$_2$ layers participate in the leading bilayer modes, whereas the leading heterostructure modes are concentrated in MoS$_2$ with little motion of hBN. The heterostructure is therefore the most physically relevant model and, after aligning the calculated and experimental ZPL positions, shows excellent agreement with the observed PL lineshape and relative PSB intensity. This agreement supports the assignment of the experimental emission to the CH$_\mathrm{S}^{-}$ defect.

\subsection{Lifetime and dipole orientation}
Beyond the PL spectra, in this section, we characterize the calculated radiative lifetimes and transition-dipole orientations of CH$_\mathrm{S}^{-}$, as summarized in Tab.~\ref{tab:properties}. The radiative lifetime indicates whether the optical transition has a relatively high or low transition rate, while the dipole orientation determines the polarization of the excitation and emission. These properties can serve as additional evidence for the defect assignment once the corresponding experimental data become available. Furthermore, the dipole orientation is also important for coupling the emitter to polarization-sensitive photonic structures, such as waveguides and optical cavities.\\
\indent The calculated lifetimes differ substantially among the four configurations. Monolayer MoS$_2$ has the shortest lifetime of 416.92~ns. In bilayer MoS$_2$, the defect position has a particularly strong effect: the lifetime is 1.15~ms for the top-bilayer configuration, compared with 28.2~$\mu$s for the middle-bilayer configuration. The MoS$_2$/hBN heterostructure has a lifetime of 16.0~$\mu$s. \\
\indent The transition-dipole orientations also vary among the configurations. Both the excitation and emission dipoles lie predominantly in the $xy$ plane, as indicated by their polar angles relative to the $z$ axis ($\theta$). Their in-plane directions, however, differ, and the misalignments between the excitation and emission dipoles are $78.78^\circ$, $56.89^\circ$, $22.96^\circ$, and $82.06^\circ$ for the monolayer, top-bilayer, middle-bilayer, and heterostructure configurations, respectively. Thus, the presence of an adjacent layer does not alter the predominantly in-plane character of the transition dipoles, although the relative orientation of the excitation and emission dipoles is sensitive to the host environment and defect position.\\
\indent Since experimental lifetimes and dipole orientations are not yet available for a direct comparison, these calculated properties provide additional predictions that could help identify the proposed CH$_\mathrm{S}^{-}$ configurations in future measurements.

\begin{table}[ht!]
\centering
\caption{Summary of the optical properties of CH$_\mathrm{S}^{-}$ in MoS$_2$. $\theta$ is the angle relative to the $z$-axis, and $\phi$ is the angle relative to the $x$-axis in the $xy$-plane. Misalignment refers to the angular difference between the excitation and emission dipoles.}
\label{tab:properties}
\resizebox{\columnwidth}{!}{%
\begin{tabular}{lcccc}
\hline
Property & Monolayer & Top bilayer & Middle bilayer & MoS$_2$/hBN \\
\hline
$Q$ (amu$^{1/2}$\AA) & 1.99 & 1.06 & 1.13 & 1.04\\
ZPL (eV) & 2.14 & 1.87 & 1.88 & 2.01 \\
ZPL (nm) & 578.90 & 661.97 & 659.75 & 616.00\\
HR factor & 10.07 & 2.12 & 2.39  & 2.70\\
DW factor & $4.23\times10^{-5}$ & 0.12 & 0.09 & 0.07 \\
Lifetime (ns) & 416.92 & $1.15\times10^{6}$ & $2.82\times10^{4}$ & $1.60\times10^{4}$ \\
$\theta_\mathrm{excitation}$ ($^\circ$) & 89.26 & 78.61 & 88.44 & 88.91 \\
$\phi_\mathrm{excitation}$ ($^\circ$) & 173.18 & 27.15 & 2.22 & 131.56\\
$\theta_\mathrm{emission}$ ($^\circ$) & 87.94 & 68.77 & 87.89 & 84.94\\
$\phi_\mathrm{emission}$ ($^\circ$) & 71.98 & 74.62 & 159.55 & 33.49\\
Misalignment ($^\circ$) & 78.78 & 56.89 & 22.96 & 82.06\\
\hline
\end{tabular}%
}
\end{table}

\section{Discussion} 
In this work, we have established a theoretical description of CH$_\mathrm{S}^{-}$ defect as the microscopic origin of the localized emission observed experimentally in MoS$_2$. Although the defect retains a similar local geometry and singlet ground state in all four considered structures, its photophysical properties depend strongly on the surrounding layers. The addition of another MoS$_2$ or hBN layer modifies the electronic transition, ZPL, PL spectrum, radiative lifetime, and transition-dipole orientation. These results demonstrate that the optical properties of CH$_\mathrm{S}^{-}$ are governed not only by its local atomic structure, but also by the surrounding layered environment.\\
\indent Despite the overall agreement between theory and experiment, two aspects require further consideration. The first concerns the electronic character of the optical transition. In our calculation, the transition in the heterostructure occurs from an occupied defect state located close to the valence-band edge to an unoccupied defect state located close to, and mixed with, the conduction-band states. The transition therefore contains contributions from CH$_\mathrm{S}^{-}$ in both the initial and final states. In contrast, the experimental emission was interpreted as an acceptor-bound exciton, A$^{-}$X, arising from recombination between a conduction-band state and an occupied acceptor-like defect state \cite{arXiv:2510.15676}. Both descriptions identify the occupied CH$_\mathrm{S}^{-}$ state as essential to the emission, but they differ in the character of the unoccupied state.\\ 
\indent Such the position and character of the unoccupied defect state may be influenced by strain and by the incomplete representation of the experimental heterostructure. Since this state lies close to the conduction-band edge, a shift in its energy or an increase in its hybridization with conduction-band states could make the calculated transition more similar to the band-to-defect transition proposed experimentally. The present model contains only one hBN layer and does not include the second hBN layer or the graphene substrate. These additional layers may modify the dielectric screening, band alignment, and hybridization between the defect and host states. This question is therefore left for future investigation with sufficient computational resources. A rigorous determination of the electron--hole composition and binding energy would further clarify this point, but this requires a $GW$-BSE calculation for the defective heterostructure, which is beyond the computational capability of the present work.\\ 
\indent The second aspect concerns the absolute ZPL value. The calculated ZPL energy of the MoS$_2$/hBN model remains slightly higher than the experimental emission peak. Our strained-monolayer calculations, which explicitly contain negatively charged CH$_\mathrm{S}^{-}$, show that the ZPL of the defect is sensitive to uniaxial and biaxial strains. It is therefore reasonable to expect that strain can also modify the ZPL in the heterostructure. Furthermore, the commensurate MoS$_2$/hBN model contains a residual strain of approximately $0.69\%$, accommodated primarily by the hBN layer. Although this strain is not applied directly to the defect-containing MoS$_2$ layer, it may still affect the CH$_\mathrm{S}^{-}$ transition indirectly by changing the interlayer interaction and electronic environment. Its effect cannot be quantified from the strained-monolayer calculations alone and may differ from the response obtained by directly straining MoS$_2$.\\ 
\indent In addition, whereas optical emission in MoS$_2$ commonly originates from excitonic recombination, the present work focuses on the defect-mediated transition associated with CH$_\mathrm{S}^{-}$ and does not explicitly investigate emission from strain-localized excitons. Spatially varying strain is known to funnel excitons toward lower-energy regions in TMDs \cite{10.1039/c7nr03537c}. Such a mechanism differs from the transition considered here. In conventional strain-funneled exciton emission, the strain-induced potential localizes host excitons. In the present system, however, the localized emission requires the electronic states introduced by CH$_\mathrm{S}^{-}$. A strain gradient could assist the localization of carriers or excitons near the defect, but it would not replace the role of the CH$_\mathrm{S}^{-}$ states in the optical transition. Our results therefore agree with the experimental assignment regarding the essential role of the defect states \cite{arXiv:2510.15676}, while the possible interplay between strain-induced exciton funneling and CH$_\mathrm{S}^{-}$-mediated emission remains to be investigated in future work.\\
\indent Despite the challenges of precisely reproducing the experimental ZPL and responsible states for the transition, the close match between the calculated and measured PL lineshapes provides strong support for the proposed CH$_\mathrm{S}^{-}$ assignment. The isolated-monolayer model produces a broad and dominant PSB that is inconsistent with the measured spectrum. In contrast, the MoS$_2$/hBN model reproduces the weak PSB and the overall experimental lineshape after alignment of the ZPL positions. In the monolayer, the stronger electron--phonon coupling is associated with a larger structural reorganization and with defect motion in one of the dominant phonon modes. In the heterostructure, the dominant modes are instead associated mainly with the MoS$_2$ host, without appreciable motion of either CH$_\mathrm{S}^{-}$ or hBN. The presence of hBN therefore modifies the vibronic response even though hBN does not participate appreciably in the dominant coupled modes.\\ 
\indent Although the bilayer MoS$_2$ models yield ZPL energies closer to the experimental peak, they do not represent the experimentally established monolayer MoS$_2$/hBN(2L)/graphene structure. Their closer numerical agreement is therefore insufficient to assign the emission to a bilayer environment. Among the models considered here, the MoS$_2$/hBN heterostructure provides the most physically relevant representation of the experimental system. The agreement in the PL lineshape and relative PSB intensity consequently provides the strongest theoretical evidence for assigning the observed emission to CH$_\mathrm{S}^{-}$.\\
\indent The calculated radiative lifetimes and transition-dipole orientations provide additional predictions for future experimental comparison. Time-resolved and polarization-resolved measurements could test the proposed assignment independently of the ZPL energy and PL lineshape. On the theoretical side, an important next step is to reduce the residual strain in the commensurate heterostructure model and determine how the transition energy and state character evolve as the lattice mismatch is reduced. Calculations including the second hBN layer and graphene would further clarify the effects of dielectric screening, interlayer coupling, and the local electrostatic environment, although such calculations remain computationally challenging at the HSE06 level.\\ 
\indent In summary, our results have established CH$_\mathrm{S}^{-}$ as a well-defined microscopic origin of localized emission in MoS$_2$. The close agreement between the calculated and experimental PL lineshapes provides strong evidence for this assignment and highlights the importance of the MoS$_2$/hBN environment compared with the isolated monolayer MoS$_2$. More broadly, this work provides microscopic guidance for identifying and engineering defect-bound optical transitions across monolayer, bilayer, and heterostructure systems. Our findings demonstrate that an adjacent layer is not merely a passive substrate, but an important factor governing the electronic, vibronic, and optical properties of defects in 2D materials.

\section{Method}
\subsection{DFT Computational details} \label{sec:method-DFT}
To theoretically investigate the CH$_\mathrm{S}^{-}$ defect in MoS$_2$, we performed first-principles calculations using the Vienna \textit{Ab initio} Simulation Package (VASP) \cite{vasp1,vasp2}. The electron-ion interaction was described within the projector augmented-wave (PAW) formalism \cite{paw,paw2}. The electronic structure was calculated using the screened hybrid Heyd-Scuseria-Ernzerhof functional (HSE06) \cite{10.1063/1.1564060}, which provides an improved description of the band gap and defect levels compared with semilocal functionals. Spin polarization was included in all calculations to properly describe the spin-dependent electronic structure and optical transitions.\\
\indent The CH$_\mathrm{S}^{-}$ defect was constructed by replacing one S atom in MoS$_2$ with a C--H complex, as illustrated in Fig.~\ref{fig:structure}. Four systems were considered: monolayer MoS$_2$, bilayer MoS$_2$ with CH$_\mathrm{S}^{-}$ at two inequivalent positions, and a MoS$_2$/hBN heterostructure. For bilayer MoS$_2$, the 2H phase with AA$'$ stacking was adopted. The two defect configurations correspond to CH$_\mathrm{S}^{-}$ located at the outer chalcogen site exposed to vacuum and at the inner chalcogen site facing the adjacent MoS$_2$ layer, as shown in Figs.~\ref{fig:structure}\textbf{b} and \textbf{c}, respectively. These configurations are hereafter referred to as the top- and middle-bilayer configurations. All remaining S sites are symmetry-equivalent to one of these two configurations. A $7\times7\times1$ supercell was employed for monolayer and bilayer MoS$_2$, containing 148 and 295 atoms after introducing CH$_\mathrm{S}^{-}$, respectively.\\
\indent For the MoS$_2$/hBN heterostructure, the lattice mismatch between MoS$_2$ and hBN was minimized by constructing a commensurate supercell composed of an $11\times11\times1$ MoS$_2$ layer and a $14\times14\times1$ hBN layer. The resulting heterostructure contains 756 atoms and has a residual strain of approximately $0.69\%$ in the hBN layer.\\
\indent In each model, CH$_\mathrm{S}^{-}$ was placed near the center of the MoS$_2$ layer. A vacuum spacing of 15~\AA\ was introduced along the out-of-plane direction to minimize interactions between periodically repeated layers. The vacuum spacing was tested using the monolayer model, for which vacuum thicknesses of 15 and 35~\AA\ yield ZPL energies of 2.1420 and 2.1419~eV, respectively. The negligible difference of 0.0001~eV confirms that the ZPL is converged with respect to the vacuum spacing. Accordingly, a vacuum spacing of 15~\AA\ was adopted for all subsequent calculations. The supercell size was examined by verifying that the defect-induced bands remain sufficiently dispersionless, indicating negligible interactions between periodically repeated defects. Brillouin-zone integrations were therefore performed using the $\Gamma$ point only. A plane-wave cutoff energy of 500~eV was employed for all calculations.\\
\indent Long-range van der Waals interactions were included using the D3 correction of Grimme \textit{et al.} \cite{10.1063/1.3382344}. This correction is necessary for describing the interlayer interactions in bilayer MoS$_2$ and the MoS$_2$/hBN heterostructure. For monolayer MoS$_2$, we verified that its inclusion has a negligible effect on the calculated defect properties. During structural optimization, the lattice parameters were fixed, while all internal atomic coordinates were relaxed until the residual force on each atom was below 0.01~eV~\AA$^{-1}$. The electronic self-consistency criterion was set to $10^{-4}$~eV.\\
\indent  The excited-state configurations were obtained using the $\Delta$SCF method, in which the electronic occupation was constrained to promote an electron into the target excited state \cite{10.1103/RevModPhys.61.689}. The ZPL energy was then calculated from the total-energy difference between the relaxed excited-state and ground-state configurations.

\subsection{PL simulation} \label{sec:method-PL}
To compare the calculated emission with the PL measurements, we simulated the PL spectrum associated with the identified optical transition using the full-phonon generating-function approach \cite{10.1088/1367-2630/16/7/073026}, as implemented in the PyPhotonics workflow \cite{10.1016/j.cpc.2021.108222}. The phonon modes were obtained from finite-displacement calculations using Phonopy \cite{phonopy-phono3py-JPCM,phonopy-phono3py-JPSJ}. All calculated vibrational modes were included in the PL simulation. The structural displacement between the relaxed ground- and excited-state geometries was projected onto each phonon mode through the configuration coordinate 
\begin{equation} q_k = \sum_{\alpha,i} \sqrt{m_\alpha} \left( R_{e,\alpha i}-R_{g,\alpha i} \right) \Delta r_{k,\alpha i},
\end{equation} 
where $\alpha$ labels the atoms, $i$ denotes the Cartesian directions, $m_\alpha$ is the mass of atom $\alpha$, $R_{g,\alpha i}$ and $R_{e,\alpha i}$ are the equilibrium coordinates of the ground and excited states, respectively, and $\Delta r_{k,\alpha i}$ is the normalized eigenvector of phonon mode $k$. The contribution of each phonon mode to the electron--phonon coupling was quantified using the partial HR factor 
\begin{equation} S_k=\frac{\omega_k q_k^2}{2\hbar}, 
\end{equation} 
where $\omega_k$ is the angular frequency of mode $k$. The total HR factor is then obtained as 
\begin{equation} S=\sum_k S_k. \end{equation} 
The corresponding HR spectral density is defined as 
\begin{equation} \mathcal{S}(\hbar\omega) = \sum_k S_k \delta \left( \hbar\omega-\hbar\omega_k \right). 
\end{equation} 
The HR spectral density resolves the total electron--phonon coupling according to the phonon energy. The peaks with the largest spectral weights identify the phonon modes with the largest partial HR factors and therefore the dominant contributions to the vibronic coupling. The DW factor, which gives the fraction of the total emission associated with the ZPL within the zero-temperature harmonic approximation, is calculated as 
\begin{equation} \mathrm{DW}=e^{-S}. 
\end{equation} 
To construct the PL spectrum, the HR spectral density was transformed into the time domain according to 
\begin{equation} \mathcal{S}(t) = \int_0^\infty \mathcal{S}(\hbar\omega) e^{-i\omega t} \,d(\hbar\omega). 
\end{equation} 
The generating function is then written as 
\begin{equation} 
G(t)= \exp\left[ \mathcal{S}(t)-\mathcal{S}(0) \right]. 
\end{equation} 
The optical spectral function is evaluated from 
\begin{equation} A(E_\mathrm{ZPL}-\hbar\omega) = \frac{1}{2\pi} \int_{-\infty}^{\infty} G(t) e^{-i\omega t-\gamma |t|} \,dt,
\end{equation} 
where $E_\mathrm{ZPL}$ is the ZPL energy and $\gamma$ is a phenomenological broadening parameter, which is set to 0.3 meV in this work. Finally, the PL spectrum is expressed as 
\begin{equation} L(\hbar\omega) = C\omega^3 A(\hbar\omega), 
\end{equation} 
where $C$ is a normalization constant and the $\omega^3$ factor accounts for the photon-frequency dependence of spontaneous emission. The resulting spectrum contains both the ZPL and the PSB. Since the calculated vibronic lineshape may depend on the supercell size, we previously benchmarked the PL spectrum of a defect in a TMD using $4\times4\times1$, $5\times5\times1$, $6\times6\times1$, and $7\times7\times1$ supercells \cite{arxiv:2605.18501}. The PL lineshape remained unchanged for supercells of $6\times6\times1$ and larger, demonstrating convergence with respect to the supercell size. Accordingly, the $7\times7\times1$ supercell employed for monolayer and bilayer MoS$_2$ is sufficient to describe the vibronic emission of CH$_\mathrm{S}^{-}$.

\subsection{Transition dipole moment and radiative lifetime}
 The transition dipole moment was computed from the momentum matrix element
\begin{equation}
\boldsymbol{\mu} =
\frac{i\hbar}{(E_f-E_i)m}
\bra{\psi_f}\mathbf{p}\ket{\psi_i},
\label{eq:dipole}
\end{equation}
using wavefunctions extracted independently from the relaxed ground- and excited-state configurations with a modified PyVaspwfc workflow \cite{pyWave,10.1088/1361-648X/ab94f4}. The dipole orientation was characterized by $\theta=\arccos(\mu_z/|\boldsymbol{\mu}|)$ and $\phi=\operatorname{atan2}(\mu_y,\mu_x)$, where $\theta$ is measured from the out-of-plane direction. The radiative lifetime was estimated from $\tau_\mathrm{R}=1/\Gamma_\mathrm{R}$, with
\begin{equation}
\Gamma_{\mathrm{R}} =
\frac{n_D e^2}{3\pi\epsilon_0\hbar^4c^3}
E_0^3 |\boldsymbol{\mu}|^2,
\end{equation}
where $E_0$ is the ZPL energy, $|\boldsymbol{\mu}|$ is the transition-dipole magnitude, and $n_D$ is the refractive index. In this work, we use $n_D=4.431$ for monolayer MoS$_2$ and the MoS$_2$/hBN heterostructure, and $n_D=4.3595$ for bilayer MoS$_2$ \cite{10.1002/adom.201900239,10.1038/s41597-023-02898-2}.

%\newpage
\section*{Data availability}
All data from this work is available from the authors upon reasonable request.

\section*{Notes}
The authors declare no competing financial interest.

\begin{acknowledgments}
This research is part of the Munich Quantum Valley, which is supported by the Bavarian state government with funds from the Hightech Agenda Bayern Plus. This work was funded by the Deutsche Forschungsgemeinschaft (DFG, German Research Foundation) under Germany's Excellence Strategy- EXC-2111-390814868 (MCQST). The authors acknowledge support from the Federal Ministry of Research, Technology and Space (BMFTR) under grant number 13N16292 (ATOMIQS). The authors gratefully acknowledge the Gauss Centre for Supercomputing e.V.\ (www.gauss-centre.eu) for funding this project by providing computing time on the GCS Supercomputer SuperMUC-NG at Leibniz Supercomputing Centre (www.lrz.de) and on its Linux-Cluster. We would like to thank Dr. Bruno Schuler and co-authors in Ref.~\cite{arXiv:2510.15676} for sharing the raw experimental data.
\end{acknowledgments}

%\clearpage

\bibliography{main}

\clearpage

\end{document}